\documentclass[9pt,twocolumn,twoside]{osajnl}

\journal{ol} 

\setboolean{shortarticle}{true}

\usepackage[detect-weight=true, detect-family=true]{siunitx}
\usepackage{txfonts}

\title{Efficient and compact source of tuneable ultrafast deep ultraviolet laser pulses at 50~kHz repetition rate}

\author[1,*]{Christian Brahms}
\author[1]{John C. Travers}

\affil[1]{School of Engineering and Physical Sciences, Heriot-Watt University, Edinburgh, EH14 4AS, UK}

\affil[*]{Corresponding author: c.brahms@hw.ac.uk}

\begin{abstract}
Deep ultraviolet (DUV) laser pulses with tuneable wavelength and very short duration are a key enabling technology for next-generation technology and ultrafast science. Their generation has been the subject of extensive experimental effort, but no technique demonstrated thus far has been able to meet all requirements in one light source. Here we demonstrate a bright, efficient, and compact source of tuneable deep ultraviolet ultrafast laser pulses based on resonant dispersive wave emission in hollow capillary fibre. In a total footprint of only $\SI{120}{\cm}\times\SI{75}{\cm}$, including the ytterbium-based drive laser, we generate pulses between \SI{208}{\nm} and \SI{363}{\nm} at \SI{50}{\kHz} repetition rate with a total efficiency of up to \SI{3.6}{\percent}. Down-scaling of the DUV generation reduces the required energy sufficiently to enable the generation of two-colour few-femtosecond DUV pulses. 
\end{abstract}

\setboolean{displaycopyright}{true}

\begin{document}

\maketitle
Laser pulses in the deep ultraviolet (DUV, \SIrange{200}{300}{\nm}) with few-femtosecond duration are a vital ingredient in next-generation technology and experiments in ultrafast science~\cite{kotsina_ultrafast_2019,teles-ferreira_ultrafast_2022,maiuri_ultrafast_2020,chergui_ultrafast_2019, kotsina_spectroscopic_2022}. Because many materials and molecular systems exhibit strong absorption features in this spectral regions, ultrafast DUV light sources enable investigation and control of new photophysical and photochemical processes. To resonantly address transitions in a wide variety of systems, wavelength tuneability is of critical importance. While ultrafast laser operation and pulse compression in the DUV have been demonstrated~\cite{nagy_generation_2009, sharp_generation_2021}, these approaches do not allow for wavelength tuneability or the generation of sufficiently short pulses. Nonlinear frequency conversion from longer-wavelength ultrafast lasers is thus the only way to generate the required radiation. However, established techniques suffer from complementary drawbacks. On the one hand, crystal-based conversion schemes can provide tuneable DUV pulses, but strong material absorption and dispersion limit the achievable lifetime and bandwidth~\cite{baum_tunable_2004, beutler_generation_2009}. On the other hand, low-order harmonic generation in gas targets driven by few-cycle infrared pulses can generate DUV pulses as short as \SI{2}{\fs}, but only at the cost of very low conversion efficiency and a fixed wavelength~\cite{graf_intense_2008,reiter_generation_2010,galli_generation_2019}. In addition, as ultrafast science moves from titanium-doped sapphire lasers at \SI{800}{\nm} towards higher-repetition-rate sources enabled by ytterbium-based lasers operating at \SI{1030}{\nm}, higher harmonic orders are required to reach the DUV which further reduces the efficiency. Resonant dispersive wave (RDW) emission in gas-filled hollow-core waveguides promises to fill this gap by generating continuously wavelength-tuneable ultrashort DUV laser pulses with high efficiency, few-femtosecond duration, and near-perfect beam quality~\cite{joly_bright_2011,mak_tunable_2013,kottig_generation_2017,travers_high-energy_2019}.

\begin{figure}[b!]
  \centering
  \includegraphics[width=3.4in]{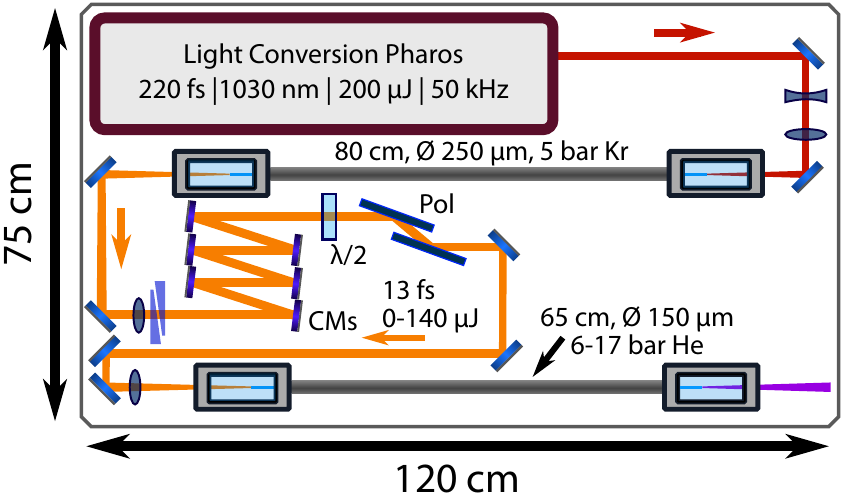}
  \caption{
  Layout of the light source. A commercial laser emits \SI{200}{\micro\joule}, 220~fs pulses at 1030~nm at a repetition rate of 50~kHz. These are compressed to 13~fs duration by spectral broadening in a first gas-filled HCF and dispersion compensation by chirped mirrors (CMs) and a wedge pair. A half-wave plate ($\lambdaup$/2) and two reflective polarisers (Pol) form a variable attenuator. DUV pulses are generated in a second HCF filled with helium.}
  \label{fig:setup}
\end{figure}

Ultraviolet RDW emission was first demonstrated in small-core antiresonant hollow fibre at low pulse energy~\cite{joly_bright_2011}, and this platform has enabled conversion to the DUV with up to \SI{2.5}{\percent} overall efficiency~\cite{kottig_generation_2017} and \SI{3}{\fs} pulse duration~\cite{brahms_direct_2019}. However, guidance resonances---spectral regions of high loss and strong dispersion, which are inherent to this type of waveguide---create gaps in the tuning range. Additionally, the small core size limits the overall pulse energy, and because inter-pulse build-up effects due to photoionisation and plasma dynamics make the use of high (\si{\mega\hertz}-scale) repetition rates difficult~\cite{koehler_long-lived_2018}, this clamps the achievable average power. Simple hollow capillary fibres (HCFs) are free from resonances and allow for significant energy scaling of RDW emission by increasing the core size~\cite{travers_high-energy_2019}. However, this approach has so far only been demonstrated using high-energy Ti:Sapphire lasers~\cite{travers_high-energy_2019,brahms_infrared_2020}. This means the overall footprint of such light sources is large, even though the length of the HCF itself can be reduced to as little as \SI{15}{\cm} by compressing the driving pulses to the few-cycle regime~\cite{brahms_high-energy_2019}. In addition, depending on such complex laser systems makes robust long-term operation and widespread adoption of HCF-based DUV sources challenging, and operation at high repetition rate (tens or hundreds of \si{\kilo\hertz}) remains out of reach. Since cutting-edge ultrafast spectoscopy experiments greatly benefit from the increased statistics offered by high-repetition-rate sources~\cite{saule_high-flux_2019}, a DUV source combining the advantages of robust ytterbium-based laser systems and RDW emission in HCF would be a very useful scientific tool.

Here we demonstrate an efficient, high-power source of ultrafast DUV laser pulses which employs RDW emission in HCF driven by a compact and robust industrialised ytterbium-based pump laser. The source operates at a repetition rate of up to \SI{50}{\kHz} and generates pulses across the DUV, with a maximum average power of over \SI{363}{\mW} at a central wavelength of \SI{233}{\nm}. This corresponds to a total conversion efficiency of \SI{3.6}{\percent}. The combination of an industrialised laser system, optimised pump-pulse compression, and the intrinsic wavelength scaling of soliton dynamics makes the source significantly more compact than previous HCF-based sources, with a total footprint of only $\SI{120}{\cm}\times\SI{75}{\cm}$ including the pump laser.

The light source is driven by a diode-pumped Yb:KGW laser amplifier (Light Conversion Pharos), which produces pulses centred at \SI{1030}{\nm} with a duration of \SI{220}{\fs} and \SI{200}{\uJ} of energy. The layout consists of two stages employing stretched HCF~\cite{nagy_flexible_2008}, as shown in Fig.~\ref{fig:setup}. The first-stage HCF serves to broaden the bandwidth of the pulses via self-phase modulation. It has a core diameter of \SI{250}{\um}, a length of \SI{80}{\cm}, and is filled with krypton at a pressure of \SI{5}{\bar}. The spectrally broadened pulses are compressed with dispersive mirrors and a wedge pair and pass through a broadband variable attenuator. We obtain up to \SI{140}{\uJ} of energy in 13 fs pulses, corresponding to 17-fold temporal compression with 70\% throughput from the laser output. The main source of energy loss is the HCF itself (81\% throughput at the fundamental limit) because the core has to be small to keep the HCF short; relaxing the length limitation would lead to even higher efficiency. DUV pulses are generated in a second stage. This consists of a smaller HCF with \SI{150}{\um} core diameter and \SI{65}{\cm} length which is filled with helium. The interplay between anomalous dispersion and the Kerr nonlinearity leads to extreme self-compression of the pulse and the subsequent generation of phase-matched resonant dispersive waves, the central wavelength of which is tuneable by simply changing the gas pressure~\cite{joly_bright_2011,travers_high-energy_2019}.

\begin{figure}
  \centering
  \includegraphics[width=3.4in]{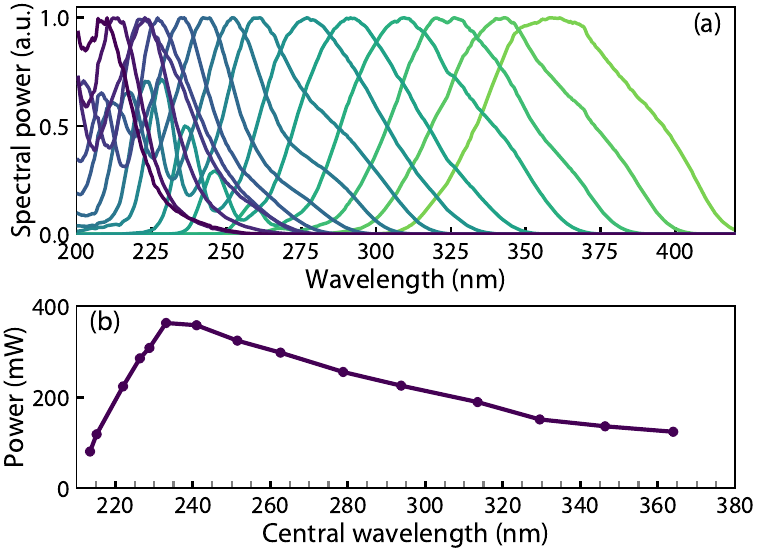}
  \caption{(a) Example spectra of DUV pulses generated in the second-stage HCF. Each curve corresponds to a different gas pressure and pump pulse energy. Longer-wavelength spectra are generated with higher pressure and less pump energy. (b) Average power at \SI{50}{\kHz} for different central wavelengths.}
  \label{fig:spectra_power}
\end{figure}

Figure~\ref{fig:spectra_power}(a) shows normalised spectra of the generated DUV pulses as the helium pressure is changed from \SIrange{6.1}{16.7}{\bar}, with longer wavelengths being generated at higher pressure. With this pressure range, the central wavelength (taken here as the first moment of the spectral power density) can be continuously tuned from \SIrange{208}{364}{\nm}. The average power of these pulses at \SI{50}{\kHz} repetition rate is shown in Fig.~\ref{fig:spectra_power}(b). It is extracted from the spectral power density measured with an integrating sphere and two fibre-coupled spectrometers which have been calibrated using NIST-traceable lamps. The highest power of \SI{363}{\mW} is generated at \SI{233}{\nm}---this corresponds to a pulse energy of \SI{7.3}{\uJ} and a total conversion efficiency from the pump laser of \SI{3.6}{\percent}. At longer wavelengths, the higher nonlinearity with higher pressure reduces the maximum energy that can be used to pump RDW emission, leading to lower average powers (\SI{123}{\mW} at \SI{364}{\nm}). At short wavelengths, the measurement is limited by the detection range of our spectrometer (200-2500~nm); it is likely that the true average power is higher, especially around the shortest central wavelength generated here. For each spectrum we extract the transform-limited pulse by taking the Fourier transform of the spectrum assuming a flat spectral phase. The bandwidth we generate supports sub-\SI{3}{\fs} pulse durations (full width at half maximum) for all central wavelengths.

\begin{figure}
  \centering
  \includegraphics[width=3.4in]{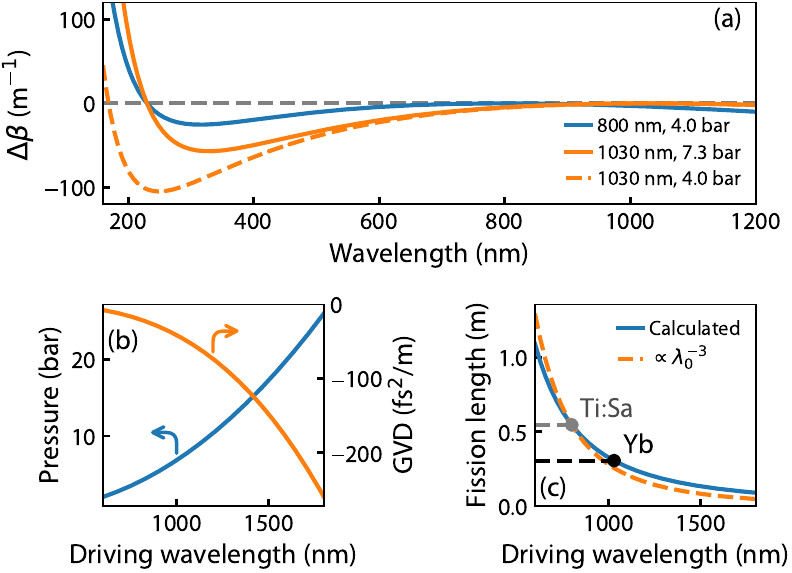}
  \caption{(a) Phase mismatch between the soliton the dispersive wave for three combinations of gas pressure and driving wavelength. The gas pressure is the same for the blue and orange dashed line. For the solid orange line, the gas pressure is adjusted to match the phase-matching point on the blue line. (b) Calculated phase-matching pressure (blue line, left axis) and total GVD (orange line, right axis) for RDW emission at \SI{230}{\nm} in \SI{150}{\um} core diameter HCF driven with \SI{13}{\fs} pulses at different central wavelengths. (c) Resulting change in the calculated fission length (blue line) and approximate scaling with $\lambda_0^{-3}$ (orange dashed line) for a driving energy of \SI{100}{\uJ}. The wavelengths and fission lengths for Ti:Sapphire and Yb-based laser systems are marked in grey and black, respectively.}
  \label{fig:scaling}
\end{figure}

The required length of HCF is determined by the length scale of soliton self-compression as encoded in the fission length~\cite{travers_high-energy_2019,brahms_high-energy_2019}. For a fixed pump wavelength and zero-dispersion wavelength (which also approximately fixes the wavelength of the dispersive wave), this scales as
\begin{equation}
  L_\mathrm{f} \propto \frac{\tau a^2}{\sqrt{I_0}}\,,
\end{equation}
where $\tau$ is the driving-pulse duration, $I_0$ its intensity, and $a$ is the HCF core radius. In order to use all of the available energy in our compressed infrared pulses, the core cannot be too small, since $I_0$ is limited by the need to avoid excessive ionisation or self-focusing~\cite{travers_high-energy_2019}. Therefore, short driving pulses are required to keep the system compact. While the driving pulses in our second stage are somewhat longer than those used for RDW emission with 800-nm lasers~\cite{travers_high-energy_2019,brahms_high-energy_2019}, this is counteracted by an additional effect---the intrinsic wavelength scaling of RDW emission. For a fixed HCF core size, the gas pressure required to phase-match RDW emission at a certain wavelength increases for longer driving wavelengths. This is illustrated in Fig.~\ref{fig:scaling}(a), where we plot the phase mismatch $\Delta\beta$ between the higher-order soliton and the dispersive wave; the phase-matching equations are available in the literature~\cite{austin_dispersive_2006,mak_tunable_2013,brahms_resonant_2020}. A helium pressure of \SI{4}{\bar} leads to phase-matching at \SI{230}{\nm} and \SI{170}{\nm} for a pump pulse at \SI{800}{\nm} and \SI{1030}{\nm}, respectively. Increasing the pressure to \SI{7.3}{\bar} for \SI{1030}{\nm} matches the RDW wavelength. The higher pressure increases the nonlinearity. At the same time, for longer wavelengths the negative contribution to the group-velocity dispersion (GVD) arising from the waveguide also becomes stronger [see Fig.~\ref{fig:scaling}(b)]. In combination, these effects cause the fission length to decrease quickly, scaling approximately as
\begin{equation}
  L_\mathrm{f} \propto \frac{\tau a^2}{\sqrt{I_0} \lambda_0^3}\,,
\end{equation}
where $\lambda_0$ is the driving wavelength. Note that this scaling rule is empirical and approximate. Because the exact behaviour is determined in part by the higher-order dispersion of the gas, it deviates from the scaling rule. Figure~\ref{fig:scaling}(c) shows a comparison between the full calculation for the fission length~\cite{travers_high-energy_2019} and applying the $\lambda_0^{-3}$ scaling rule starting with the value at \SI{800}{\nm}.  For constant pulse duration, the fission length is reduced by nearly half when moving from \SI{800}{\nm} to \SI{1030}{\nm}. It is this scaling which makes our extremely compact source possible.

\begin{figure}
  \centering
  \includegraphics[width=3.4in]{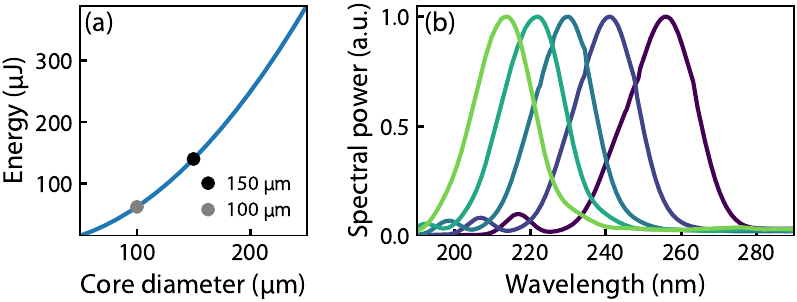}
  \caption{(a) Calculated energy required for RDW emission for different HCF core sizes, assuming the full compressor output of \SI{140}{\uJ} is required for a core diameter of \SI{150}{\um}. (b) RDW spectra generated in a second-stage HCF with \SI{100}{\um} core diameter for helium pressures between \SI{13.5}{\bar} and \SI{19}{\bar}.}
  \label{fig:downscaled}
\end{figure}

\begin{figure}
  \centering
  \includegraphics[width=3.4in]{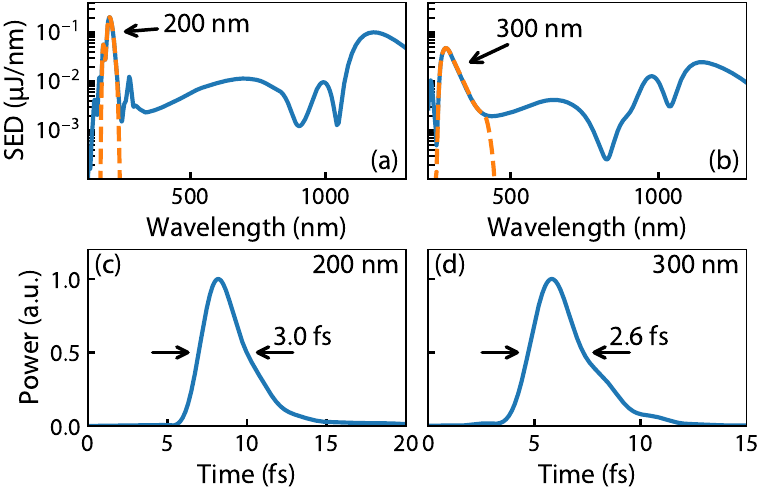}
  \caption{
  Simulations of two-colour ultrafast DUV pulse generation. (a) Power spectrum generated in a \SI{14}{\cm} long, \SI{100}{\um} core diameter HCF filled with a pressure gradient from \SI{25}{\bar} of helium to vacuum and pumped with \SI{13}{\fs} pulses with \SI{55}{\micro\joule} of energy. The orange dashed line shows the filtered spectrum of the DUV pulse. (b) Same as (a) but for \SI{20}{\cm} HCF length, \SI{1.8}{\bar} of argon gas and \SI{28.5}{\micro\joule} energy. (c), (d) DUV pulse profiles contained in the filtered spectra shown in (a) and (b), respectively.}
  \label{fig:simulations}
\end{figure}

Many ultrafast spectroscopy experiments benefit from pumping and probing with pulses at different wavelengths. By exploiting the energy scaling laws for soliton dynamics and general nonlinear optics in gases~\cite{travers_high-energy_2019,heyl_scale-invariant_2016}, we can reduce the energy required in the second HCF stage and therefore provide enough energy to pump two independent UV generation stages, enabling two-colour experiments. Scaling the transverse dimension (here, the core diameter of the HCF) by a factor $\eta$ scales the overall energy and longitudinal dimension (here, HCF length) by $\eta^2$, and the gas density by $1/\eta^2$~\cite{heyl_scale-invariant_2016}. Fig.~\ref{fig:downscaled}(a) shows that moving from a core diameter of \SI{150}{\um} to \SI{100}{\um} reduces the overall energy scale of the process by a factor of $(150/100)^2 = 2.25$. Even considering additional losses, splitting the beam evenly into two arms will thus enable DUV generation in each arm. To demonstrate this potential, we replace the second stage with \SI{40}{\cm} of \SI{100}{\um} core diameter HCF. Figure \ref{fig:downscaled}(b) shows a series of spectra generated in this smaller HCF with helium pressures between \SI{13.5}{\bar} and \SI{19}{\bar}. The pulse energy required for RDW emission, as measured just before the coupling lens, is \SI{61}{\uJ} at the longest RDW wavelength and \SI{93}{\uJ} at the shortest. The scaling laws predict that less than half of the total compressor output (\SI{140}{\uJ}) should be required to cover the same tuning range as the larger HCF. There are two reasons why we do not observe this in the experiment. Firstly, the coupling efficiency to this smaller HCF is approximately \SI{15}{\percent} lower due to increased aberrations in the focusing (for the same reason, the spectra in Fig.~\ref{fig:downscaled}(b) are collected at \SI{10}{\kilo\hertz} to avoid damaging the HCF). Secondly, the scaling laws do not account for the increased propagation losses for smaller cores~\cite{brahms_high-energy_2019}.

In Fig.~\ref{fig:simulations}, we show numerical simulations of an optimised two-colour DUV generation system. Our simulations are based on solving the multi-mode unidirectional pulse propagation equation and take into account the HCF dispersion and loss, the Kerr effect, and photoionisation and plasma dynamics~\cite{travers_high-energy_2019, brahms_lunajl_2021}. We consider \SI{13}{\fs} pump pulses and \SI{100}{\micro\meter} core diameter HCFs filled with decreasing pressure gradients, which enables the direct delivery of compressed DUV pulses to a vacuum chamber~\cite{brahms_direct_2019, brahms_resonant_2020, kotsina_spectroscopic_2022}. In Fig.~\ref{fig:simulations}(a) we show the power spectrum generated when the HCF is \SI{14}{\cm} long, its entrance is filled with \SI{25}{\bar} of helium, and the pump energy is \SI{55}{\micro\joule}. This results in the generation of a broadband resonant dispersive wave at \SI{200}{\nm}. While the gas pressure in this case is high, suitable differential pumping schemes would allow for high vacuum to be maintained in the rest of the system. Figure \ref{fig:simulations}(b) shows the power spectrum for \SI{20}{\cm} HCF length, \SI{1.8}{\bar} of argon fill pressure, and \SI{28.5}{\micro\joule} pump energy. Here the resonant dispersive wave is centred at \SI{300}{\nm}. The longer RDW wavelength allows for the use of argon and lower gas pressure in this case: because the pressure is higher for a given gas species, the gas nonlinearity is stronger, and hence the required pulse energy is lower; this is the same effect which reduces the power generated at longer wavelengths in our experiments. The lower intensity (\SI{5e13}{\watt\per\square\cm} as compared to \SI{e14}{\watt\per\square\cm} for the RDW at \SI{200}{\nm}) reduces photoionisation effects considerably and hence allows for the use of a gas with lower ionisation potential. The RDW pulses are extracted by bandpass filtering, with the filtered spectrum shown as orange dashed lines in Fig.~\ref{fig:simulations}(a) and (b). Fig.~\ref{fig:simulations}(c) and (d) show the filtered pulses. In both cases, extremely short pulses are generated, with a FWHM duration of \SI{3}{\fs} at \SI{200}{\nm} and \SI{2.6}{\fs} at \SI{300}{\nm}. An optimised system would hence allow for two-colour time-resolved DUV spectroscopy with few-femtosecond time resolution at high repetition rate.

In summary, we have demonstrated the generation of high-energy ultrafast DUV laser pulses at high repetition rate in a small footprint. Energy down-scaling of the DUV generation stage brings two-colour few-femtosecond pulse generation within reach. Our results mark the first time the numerous advantages of HCF-based DUV sources have been harnessed using an industrialised ytterbium-based laser system instead of a titanium-doped sapphire laser. This enables operation at high repetition rate and average power with excellent total conversion efficiency, while the small footprint of the laser itself in combination with the intrinsic wavelength scaling of soliton dynamics keeps the overall device compact. Tuneable ultrafast DUV pulses hold great promise for both science and technology. The compact, bright and efficient laser source we have demonstrated here will broaden access to new applications in these fields and beyond.

\begin{backmatter}
\bmsection{Funding} This work was funded by the European Research Council under the European Union Horizon 2020 Research and Innovation program: Starting Grant agreement HISOL, No.~679649; Proof of Concept Grant agreement ULIGHT, No.~899900; Consolidator Grant agreement XSOL, No.~101001534, and by the United Kingdom Engineering and Physical Sciences Research Council: Grant agreement EP/T020903/1. This project was supported by the Royal Academy of Engineering under the Research Fellowship Programme, grant agreement RF/202122/21/133.


\bmsection{Disclosures} The authors declare no conflicts of interest.

\bmsection{Data availability}
The data that support the findings of this study are available from the corresponding author upon reasonable request.

\end{backmatter}

\bibliography{bibliography}

\bibliographyfullrefs{bibliography}

\end{document}